\documentclass[showpacs,prb,manuscript]{revtex4}
\usepackage{amssymb}
\usepackage{graphicx}

\begin{document}

\title{Electron spin relaxation in semiconducting carbon nanotubes: the role
of hyperfine interaction}
\author{Y. G. Semenov, K. W. Kim, and G. J. Iafrate}
\affiliation{Department of Electrical and Computer Engineering\\
North Carolina State University, Raleigh, NC 27695-7911}
\pacs{85.35.Kt, 85.75.-d}

\begin{abstract}
A theory of electron spin relaxation in semiconducting carbon nanotubes is
developed based on the hyperfine interaction with disordered nuclei spins
I=1/2 of $^{13}$C isotopes. It is shown that strong radial confinement of
electrons enhances the electron-nuclear overlap and subsequently electron
spin relaxation (via the hyperfine interaction) in the carbon nanotubes. The
analysis also reveals an unusual temperature dependence of longitudinal
(spin-flip) and transversal (dephasing) relaxation times: the relaxation
becomes weaker with the increasing temperature as a consequence of the
particularities in the electron density of states inherent in
one-dimensional structures. Numerical estimations indicate relatively high
efficiency of this relaxation mechanism compared to the similar processes in
bulk diamond. However, the anticipated spin relaxation time of the order of
1 s in CNTs is still much longer than those found in conventional
semiconductor structures.
\end{abstract}

\maketitle


Due to their unique electrical properties, carbon nanotubes (CNTs) are
considered to be the ultimate structure for continued \textquotedblleft
scaling\textquotedblright\ beyond the end of the semiconductor
microelectronics roadmap.~\cite{Dekker} Moreover, the unique electrical
properties of CNTs are enhanced by the equally unique structural properties.
This combination assures the development of CNTs for important applications
and has largely been the focus of attention to date (see Refs.~%
\onlinecite{Dresselhaus} and \onlinecite{Ando05} as well as the references
therein). Recently, however, the researchers are beginning to explore other
important advantages that the CNTs can offer. For example, the CNTs with
naturally low or no impurity incorporation allow, in addition to the more
conventional scaled transistor application, the injection and use of
electrons with polarized spin~\cite{Tsukagoshi99,Sahoo05} as an added
variable for computation. Thus, CNTs are an ideal medium for the development
of the emerging field of spintronics.~\cite{Meh,Yang} Further, the
anticipated long spin relaxation times allow coherent manipulation of
electron spin states at an elevated temperature, opening a significant
opportunity for spin-based quantum information processing. Clearly, spin
dependent properties of CNTs warrant a comprehensive investigation from the
point of view of fundamental physics (see, for example, Refs.~%
\onlinecite{Martino02} and \onlinecite{Martino04}) and practical
applications.

The objective of the present paper is to theoretically investigate the
electron spin relaxation properties in the CNTs, a crucial piece of
information for any spin related phenomena. Specifically, we consider the
electron hyperfine interaction (HFI) with nuclear spins $I=1/2$ of $^{13}$C
isotopes (with the natural abundance of $1.10\%$). The HFI is thought to be
one of the most important spin relaxation processes in the CNTs; strong
radial confinement of electrons in the CNTs enhances electron-nuclear
overlap and subsequently the hyperfine interaction compared to the bulk
crystals. On the other hand, the mechanisms related to spin-orbital
interaction are expected to be extremely weak in CNT.\cite{AndoSOI} In the
following analysis, the main emphasis will be on the single-walled
semiconducting nanotubes.

The property of our interest is the longitudinal (T$_{1}$) and transversal (T%
$_{2}$) spin relaxation time of an electron with the radius vector $%
\overrightarrow{r}$ and spin $\overrightarrow{S}$ in a CNT. The governing
Hamiltonian caused by the Fermi contact HFI with $N$ nuclear spins $%
\overrightarrow{I}^{j}$ located at lattice sites $\overrightarrow{R}_{j}$
can be expressed as
\begin{equation}
H_{hf}=\Omega _{0}a_{hf}\sum_{j=1}^{N}\overrightarrow{S}\overrightarrow{I}%
^{j}\delta (\overrightarrow{r}-\overrightarrow{R}_{j})\equiv \overrightarrow{%
\Theta }\overrightarrow{S},  \label{eq1}
\end{equation}%
where the HFI constant $a_{hf}$ and the area of the graphene sheet $\Omega
_{0}$ are normalized per carbon atom. As indicated, this Hamiltonian $H_{hf}$
can also be expressed in terms of the fluctuating field operator $%
\overrightarrow{\Theta }$ that mediates spin relaxation.

To proceed further, $\overrightarrow{\Theta }$ must be expressed in terms of
electronic Bloch states of the relevant energy bands. In an effective mass
approximation, the eigenstates for the conduction bands in the vicinity of
the $K$ point take the form~\cite{Ando05,AjikiAndo}
\begin{equation}
\left\vert k\right\rangle =\frac{1}{\sqrt{2A_{0}L}}\left(
\begin{array}{c}
B_{\nu ,n}(k) \\
1%
\end{array}%
\right) e^{i[\text{\ae }_{\nu ,n}\xi +k\eta ]},  \label{eq5}
\end{equation}%
where $A_{0}$ denotes the length of the CNT, $\vec{L}$ ($=n_{1}%
\overrightarrow{b}_{1}+n_{2}\overrightarrow{b}_{2}$) the chiral vector in
terms of primitive translation vectors $\overrightarrow{b}_{1}$, $%
\overrightarrow{b}_{2}$ and integers $n_{1}$, $n_{2}$, and $B_{\nu ,n}(k)=[$%
\ae $_{\nu ,n}-ik]/\sqrt{\text{\ae }_{\nu ,n}{}^{2}+k^{2}}$ with \ae $_{\nu
,n}=(2\pi /L)(n-\nu /3)$; the quantum number $n=0,\pm 1,\pm 2,...$
distinguishes the energy bands, while $\nu $ takes one of the three integers
$-1,0,1$ that makes $(n_{1}-n_{2}-\nu )$ an integer multiple of $3$. As
shown in Fig.~1, $\xi $ and $\eta $ represent the coordinates for the axes
directed along $\overrightarrow{L}$ (i.e., the circumference) and the CNT
(i.e., $\overrightarrow{M}$), respectively. The eigenstates $\left\vert
k^{\prime }\right\rangle $ for the $K^{\prime }$ valley can be readily
obtained from Eq.~(\ref{eq5}) by substituting $B_{\nu ,n}\rightarrow B_{-\nu
,n}^{\ast }$, and $k\rightarrow k_{\nu ,n}^{\prime }$. The wave vectors $k$
and $k^{\prime }$ are determined from the $K$ and $K^{\prime }$ points of
the Brillouin zone, respectively.

The corresponding dispersion relation for the $\left\vert k\right\rangle $
states reads
\begin{equation}
\varepsilon _{n,k}=\gamma \sqrt{\text{\ae }_{\nu ,n}{}^{2}+k^{2}},
\label{eq6}
\end{equation}%
where $\gamma $ is a transfer matrix element. Assuming that only the lowest
conduction band is occupied by electrons in a semiconducting CNT with $\nu
=+1$ or $-1$, we restrict our consideration to the $n=0$ case at a given
temperature $T$. Then, Eq.~(\ref{eq6}) in the vicinity of the $K$ point can
be approximated as
\begin{equation}
\varepsilon _{k}=\frac{E_{g}}{2}+\frac{\hbar ^{2}k^{2}}{2m^{\ast }}
\label{eq7}
\end{equation}%
with an effective mass $m^{\ast }=2\pi \hbar ^{2}/3L\gamma $ and
the band gap $E_{g}=4\pi \gamma /3L$. In the $K^{\prime }$ valley,
a similar dispersion relation can be obtained when $k$ is
substituted by $k^{\prime }$. Although it is known that the
external magnetic field $\overrightarrow{B}$ modifies the CNT
electronic states, this effect is neglected as the relevant
parameter $(d_{t}/2a_{H})^{2}$ (where $d_{t}=|\vec{L}|/\pi $ is
the CNT diameter and $a_{H}=\sqrt{c\hbar /eB}$ the magnetic
length) is practically
very small.\cite{Ando05,AjikiAndo} Hence, we only consider the influence of $%
B$ on electron spin states through the Zeeman energy $\hbar \omega \sigma $;
$\sigma =\pm 1/2$ is the spin projection on the $\overrightarrow{B}$
direction.

Utilizing the expressions given above, we can represent the fluctuating
field operator in a second-quantized form in terms of the electron
creation-annihilation operators $a_{k,\sigma }^{\dag }$ and $a_{k,\sigma }$,
\begin{equation}
\Theta _{\mu }=\frac{a_{NT}}{A_{0}}\sum_{k_{1},k_{2},\sigma
}\sum_{j=1}^{N}e^{i(k_{1}-k_{2})\eta _{j}}I_{\mu }^{j}a_{k_{1},\sigma
}^{\dag }a_{k_{2},\sigma },  \label{eq8}
\end{equation}%
where $\mu $ denotes the coordinate for the spin states; by
convention, the direction of the magnetic field
$\overrightarrow{B}$ is chosen as the $z$ axis (quantization axis)
and two transversal directions as $x$ and $y$ ($\mu =x,y,z$). In
addition, $a_{NT}=a_{hf}\Omega _{0}/L$ and $\eta _{j}$ is the
location of the $j$-th nuclear spin on the CNT axis. As $k_{1}$
and $k_{2}$ are any two states in the Brillouin zone,
Eq.~(\ref{eq8}) accounts for the effects of both intra- and
inter-valley electron scattering on the nuclear spins.

Let us now consider the spin evolution caused by arbitrary random
fluctuations $\Theta _{\mu }(t)$. The time dependence of the mean spin value
$\vec{s}$ can be described by the quantum kinetic equation provided the spin
relaxation times $T_{1}$ and $T_{2}$ are much longer than the correlation
time of the thermal bath:~\cite{Sem03}
\begin{equation}
\frac{d}{dt}\vec{s}(t)=\vec{\omega}\times \vec{s}(t)-\mathbf{\Gamma }\left[
\vec{s}(t)-\vec{s}_{0}\right] ,  \label{eq0}
\end{equation}%
where $\vec{\omega}=\omega \overrightarrow{B}/|\overrightarrow{B}|$ if the $g
$-factor anisotropy is ignored and the electron spin polarization at thermal
equilibrium $\vec{s}_{0}$ is given as $-{\frac{1}{2}}\widehat{z}\tanh (\hbar
\omega /2k_{B}T)$ ($k_{B}$ the Boltzmann constant). Finally, the matrix $%
\mathbf{\Gamma }$ of the relaxation coefficients can be reduced to the
Bloch-Redfield diagonal form with a leading diagonal composed of matrix
elements $\Gamma _{xx}=T_{2}^{-1}$, $\Gamma _{yy}=T_{2}^{-1}$, and $\Gamma
_{zz}=T_{1}^{-1}$:
\begin{eqnarray}
T_{1}^{-1} &=&2\pi n(\omega )\gamma _{xx}(\omega ),  \label{eq2} \\
T_{2}^{-1} &=&\pi \lbrack \gamma _{zz}(0)+n(\omega )\gamma _{xx}(\omega )],
\label{eq3}
\end{eqnarray}%
where $n(\omega )=(1+e^{-\hbar \omega /k_{B}T})/2$ and $\gamma _{\mu \mu
}(\omega )$ is the Fourier transformed correlation function of the operator $%
\Theta _{\mu }$,
\begin{equation}
\gamma _{\mu \mu }(\omega )=\frac{1}{2\pi \hbar ^{2}}\int_{-\infty }^{\infty
}\left\langle \Theta _{\mu }(\tau )\Theta _{\mu }\right\rangle e^{i\omega
\tau }d\tau .  \label{eq4}
\end{equation}%
Hence, evaluation of the longitudinal $T_{1}$ and the transversal $T_{2}$
relaxation times can be reduced to finding the relevant $\gamma _{\mu \mu }$%
. In Eq.~(\ref{eq4}), $\Theta _{\mu }\left( \tau \right) =\exp (iH_{d}\tau
/\hbar )\Theta _{\mu }\exp (-iH_{d}\tau /\hbar )$, $\left\langle \ldots
\right\rangle =Tr\{e^{-H_{d}/k_{B}T}\ldots \}/Tre^{-H_{d}/k_{B}T}$, where $%
H_{d}$ is the Hamiltonian of the thermal bath. In our case, it takes the form%
\begin{equation}
H_{d}=\sum\limits_{\mathbf{k},\sigma }\varepsilon _{\mathbf{k}}a_{\mathbf{k}%
,\sigma }^{\dagger }a_{\mathbf{k},\sigma }+\sum_{j}\hbar \omega
_{n}I_{Z}^{j}.  \label{eq3b}
\end{equation}%
The first term of Eq.~(\ref{eq3b}) represents the kinetic energy of the
electron, which is basically the electron Hamiltonian after the Zeeman
energy $\sum\limits_{\mathbf{k},\sigma }\hbar \omega \sigma a_{\mathbf{k}%
,\sigma }^{\dagger }a_{\mathbf{k},\sigma }$ is removed; as defined earlier, $%
a_{\mathbf{k},\sigma }^{\dagger }$ and $a_{\mathbf{k},\sigma }$ are the
creation and annihilation operators of an electron with energy $\varepsilon
_{\mathbf{k}}$ [Eq.~(\ref{eq7})] and $\sigma $ is the electron spin quantum
number. The second term accounts for the magnetic energy due to the nuclear
spin splitting $\hbar \omega _{n}$ in a magnetic field.

As the electron momentum relaxation time $\tau _{k}$ is expected to be
shorter than the spin relaxation time, the correlation functions can be
found from Eq.~(\ref{eq4}) in terms of $\delta $-functions reflecting
conservation of energy, when the average electron kinetic energy $%
\left\langle \varepsilon _{\mathbf{k}}\right\rangle \approx k_{B}T$ is much
larger than the energy broadening $\Gamma $ of the order of $h\tau _{k}^{-1}$
(i.e., $k_{B}T\gg h\tau _{k}^{-1})$. To further simply the formulation, the
nuclear spin operator $I_{x}$ contained in the fluctuating field operator $%
\Theta _{x}$ [Eq.~(\ref{eq8})] is conveniently split into two parts $%
I_{x}=(I_{+}+I_{-})/2$ with the raising and lowering operators $I_{\pm
}=I_{x}\pm iI_{y}$; correspondingly, $\Theta _{\pm }$ is defined from $%
\Theta _{x}=(\Theta _{+}+\Theta _{-})/2$ as a formal substitution for index $%
\mu $. 
Then, by averaging $e^{i(k_{1}-k_{2})\eta _{j}}$ over the random
distribution of $N$ nuclear isotopes $^{13}$C, the Fourier transformation $%
\gamma _{\pm \mp }\left( \omega \right) $ of the correlation function $%
\left\langle \Theta _{\pm }(\tau )\Theta _{\mp }\right\rangle $ gives
\begin{equation}
\gamma _{\pm \mp }\left( \omega \right) =2N\frac{a_{NT}^{2}}{\hbar A_{0}^{2}}%
\left\langle I_{\pm }I_{\mp }\right\rangle \sum\limits_{k,k,\sigma
}f_{k,\sigma }(1-f_{k^{\prime },\sigma })\delta \left( \pm \hbar \omega
_{n}+\varepsilon _{k}-\varepsilon _{k^{\prime }}+\hbar \omega \right) .
\label{eq25}
\end{equation}%
The distribution function $f_{k,\sigma }=\left\langle a_{k,\sigma }^{\dag
}a_{k,\sigma }\right\rangle $ for non-degenerate electrons is $%
e^{-u_{k,\sigma }/k_{B}T}/\sum_{k,\sigma }e^{-u_{k,\sigma }/k_{B}T}$, where $%
u_{k,\sigma }=u_{k}+\hbar \omega \sigma $, $u_{k}=\varepsilon _{k}-E_{g}/2$.
Since $\gamma _{++}(\omega )=\gamma _{--}(\omega )=0$ from $\left\langle
I_{+}I_{+}\right\rangle =\left\langle I_{-}I_{-}\right\rangle =0$, Eq.~(\ref%
{eq25}) allows one to find $\gamma _{xx}(\omega )=\left[ \gamma _{+-}(\omega
)+\gamma _{-+}(\omega )\right] /4$ as well as $n(\omega )\gamma _{xx}(\omega
)$ in the form
\begin{equation}
n(\omega )\gamma _{xx}(\omega )=\frac{1}{8}\left[ \gamma _{+-}(\omega
)+\gamma _{-+}(\omega )+\gamma _{+-}(-\omega )+\gamma _{-+}(-\omega )\right]
.  \label{eq26a}
\end{equation}%
Using Eqs.~(\ref{eq25}) and (\ref{eq26a}) and identity $\left\langle I_{\pm
}I_{\mp }\right\rangle =\left\langle I_{x}^{2}\right\rangle +\left\langle
I_{y}^{2}\right\rangle \pm \left\langle I_{z}\right\rangle \cong
2\left\langle I_{x}^{2}\right\rangle $, one can derive relaxation parameters
in Eqs.~(\ref{eq2}) and (\ref{eq3}). Under the assumption that the nuclear
spin splitting $\omega _{n}$ is negligible compared to $\omega $, it takes
the form
\begin{equation}
\pi n(\omega )\gamma _{xx}(\omega )=N\frac{a_{NT}^{2}}{\hbar A_{0}^{2}}%
\sum_{k,k^{\prime },\sigma }\left\langle I_{x}^{2}\right\rangle \left[
f(1-f^{\prime })+f^{\prime }(1-f)\right] \left[ \delta \left( \hbar \omega
+\varepsilon -\varepsilon ^{\prime }\right) \right] ,  \label{eq27a}
\end{equation}%
where $\varepsilon =\varepsilon _{k,\sigma }$, $\varepsilon ^{\prime
}=\varepsilon _{k^{\prime },\sigma }$. Applying inequalities $f=f_{k,\sigma
}\ll 1$, $f^{\prime }=f_{k^{\prime },\sigma }\ll 1$, Eq. (\ref{eq27a}) for
non-degenerate electrons reduces to
\begin{equation}
\pi n(\omega )\gamma _{xx}(\omega )=2N\frac{a_{NT}^{2}}{\hbar A_{0}^{2}}%
\left\langle I_{x}^{2}\right\rangle \sum_{k,k^{\prime },\sigma }f_{k,\sigma
}\delta (\hbar \omega +\varepsilon _{k}-\varepsilon _{k^{\prime }}).
\label{eq9}
\end{equation}%
Similarly, we find
\begin{equation}
\pi \gamma _{zz}(0)=2N\frac{a_{NT}^{2}}{\hbar A_{0}^{2}}\left\langle
I_{z}^{2}\right\rangle \sum_{k,k\mathbf{^{\prime }},\sigma }f_{k,\sigma
}\delta (\varepsilon _{k}-\varepsilon _{k^{\prime }}).  \label{eq32}
\end{equation}%
Note that in the case of $I=1/2$, $I_{\mu }^{2}=\frac{1}{2}\widehat{\mathbf{1%
}}$ that leads to $\left\langle I_{x}^{2}\right\rangle =\left\langle
I_{z}^{2}\right\rangle =1/4$ ($\widehat{\mathbf{1}}$ is the unity matrix).

One can see that the contribution of the elastic scattering that does not
involve electron and nuclear spin flip-flop [Eq.~(\ref{eq32})] differs from
that of the inelastic process [Eq.~(\ref{eq9})] by $\hbar \omega $ in the
argument of the $\delta $-function. Subsequently, we focus on the
calculation of $\gamma _{xx}(\omega )$ that covers the case of Eq.~(\ref%
{eq32}) in the limit $\omega \rightarrow 0$. The sum over the wave vectors
in Eqs.~(\ref{eq9}) and (\ref{eq32}) can be calculated by integrating the
energy $u=\varepsilon _{k}-E_{g}/2$ with the density of states $D(u)$. In
the vicinity of each valley,
\begin{equation}
D(u)=\frac{A_{0}}{\pi \hbar }\sqrt{\frac{2m^{\ast }}{u}}  \label{eq10}
\end{equation}%
that leads to the electron distribution function in the form%
\begin{equation}
\sum_{\sigma }f_{k,\sigma }=\frac{\hbar }{2A_{0}}\sqrt{\frac{2\pi }{m^{\ast
}k_{B}T}}e^{-u_{k}/k_{B}T}.  \label{eq11}
\end{equation}

It can be shown that the double summation over $k$ and $k^{\prime }$ can be
reduced to the summation over a single valley by multiplying the result by
the valley degeneracy $l_{v}=2$. Straightforward calculation of the
integrals with the density of states $D(u)$ under the condition $\hbar
\omega \ll k_{B}T$ results in
\begin{equation}
n(\omega )\gamma _{xx}(\omega )=\frac{a_{NT}^{2}l_{v}\sqrt{m^{\ast }}}{\pi
\hbar ^{2}\sqrt{2\pi k_{B}T}}\frac{N}{A_{0}}\ln \frac{k_{B}T}{\hbar \omega }.
\label{eq12}
\end{equation}%
In the limit $\omega \rightarrow 0$, Eq.~(\ref{eq12}) reveals a logarithmic
singularity. This situation is not only typical for one-dimensional systems
but also known in the galvanomagnetic effect in bulk crystals. A standard
recipe for removing such a divergency consists of taking into account
broadening of the energy levels $\Gamma $ due to the electron scattering
processes discussed earlier. Therefore, as soon as $\hbar \omega $ becomes
smaller than this broadening factor $\Gamma $, the magnetic field dependence
becomes saturated at $\ln (k_{B}T/\Gamma )$ instead of $\ln (k_{B}T/\hbar
\omega )$ in Eq.~(\ref{eq12}). Moreover, one must also take into account the
finite length $A_{0}$ of an actual CNT. In such a case, the electron energy
cannot be less than $\Delta \varepsilon \approx \hbar ^{2}/m^{\ast
}A_{0}^{2} $, which substitutes $\Gamma $ if $\Delta \varepsilon >\Gamma $.
In the following, we assume that the effect of finite $\Delta \varepsilon $
is included in the parameter $\Gamma $. Note that a similar restriction on
the bottom limit of the electron energy $u$ would be applied to $D(u)$ in
Eq.~(\ref{eq10}). Therefore the condition for the validity of Eq.~(\ref{eq12}%
) will be satisfied by $\max \{\Gamma ,\hbar \omega \} \ll k_{B}T$.

In a manner similar to that discuss above, we can find $\gamma _{zz}(0)$,
which looks like Eq.~(\ref{eq12}) with $\hbar \omega \rightarrow \Gamma $.
The final expressions for relaxation times [Eqs.~(\ref{eq2}) and (\ref{eq3}%
)] take the form%
\begin{eqnarray}
T_{1}^{-1} &=&\tau _{hf}^{-1}\ln \frac{k_{B}T}{\max \{\hbar \omega ,\Gamma \}%
},  \label{eq13} \\
T_{2}^{-1} &=&\frac{\tau _{hf}^{-1}}{2}\left( \ln \frac{k_{B}T}{\Gamma }+\ln
\frac{k_{B}T}{\max \{\hbar \omega ,\Gamma \}}\right) ,  \label{eq14}
\end{eqnarray}%
where the essential part, which determines the order of magnitude of the
spin relaxation, can be expressed in terms of the fundamental CNT parameters
\begin{equation}
\tau _{hf}^{-1}=\frac{2l_{v}xa_{hf}^{2}\Omega _{0}}{\hbar \sqrt{3\gamma
L^{3}k_{B}T}}.  \label{eq15}
\end{equation}

Equations~(\ref{eq13}), (\ref{eq14}) and (\ref{eq15}) exhibit an unusual
temperature dependence for the spin relaxation rate; in contrast to the
three-dimensional case, decreasing $T$ enhances spin relaxation. Apparently,
this effect stems from the property of the one-dimensional density of
states, which increases as $u$ decreases to $\Gamma $. Equation~(\ref{eq15})
also shows that the geometric properties of different CNTs manifest itself
only via the length of the chirality vector as a factor $L^{-3/2}$ ($%
L=\left\vert \overrightarrow{b}_{1}\right\vert \sqrt{%
n_{1}^{2}+n_{2}^{2}+n_{1}n_{2}}$). Hence, the relaxation rates for a variety
of semiconducting CNTs can be readily compared by using this scaling rule.
As for the magnetic field dependence, it appears in Eqs.~(\ref{eq13}) and (%
\ref{eq14}) as the parameter $ \hbar \omega $, which interplays with $\Gamma
$. When $\hbar \omega > \Gamma $, the calculation predicts gradual reduction
of the spin relaxation rate as $B$ increases. On the other hand, no magnetic
field influence can be expected once $\hbar \omega$ drops below $\Gamma $.

As an example, we consider a zigzag CNT with $(n_{1},n_{2})=(8,0)$ and
assume $\Gamma =1$~$\mu $eV. Other parameters are known to be: $x=0.011$, $%
\Omega _{0}=\sqrt{3}b^{2}/4$, $b=0.249$ nm, $\gamma =\gamma _{0}\Omega
_{0}/b $, $\gamma _{0}=3.013$~eV.~\cite{Dresselhaus} The HFI constant for $%
^{13}$C was estimated in Ref.~\onlinecite{Barone}: $a_{hf}/2\pi \hbar =22.5$%
~MHz. Figure 2 presents the calculated relaxation rates $T_{1}^{-1}$ and $%
T_{2}^{-1}$ as a function of temperature at various magnetic field
strengths. Clearly, spin relaxation becomes slower with the increasing
temperature as discussed above. $T_{1}$ is always longer than $T_{2}$ with
the exception of the zero-field case, where the longitudinal and transversal
relaxations are indistinguishable. Both relaxation rates also show gradual
decrease as $B$ becomes larger. With the relaxation time of about 1~s, these
characteristics are readily observable by experiments.

In order to consider the effect of radial confinement, we calculate the
spin-flip rate $W_{d}$ in bulk diamond. In general, $W_{d}=n\overline{v}%
\sigma _{sf}$, where $n=2x/\Omega _{d}$ is the nuclear spin concentration ($%
\Omega _{d}$ is the unit cell volume of diamond), $\overline{v}=\sqrt{%
8k_{B}T/\pi m}$ the mean electron velocity at temperature $T$, and $m=\sqrt[3%
]{m_{\shortparallel }m_{\perp }^{2}}$ the density of states effective mass. $%
m_{\shortparallel }$ and $m_{\perp }$ are the longitudinal and transversal
masses in each of diamond $X$ valleys with six fold degeneracy (i.e., $%
l_{v}=6$). The spin-flip cross section for electron scattering with a
localized spin moment calculated in the first Born approximation is known to
be $\sigma _{sf}=\frac{2}{3\pi \hbar ^{4}}I(I+1)l_{v}a_{hf}^{2}\Omega
_{d}^{2}m^{2}$ [see Ref.~\onlinecite{Deigen}]. Taking into account that $%
\Omega _{d}=5.67\times 10^{-24}$ cm$^{3}$, $m_{\shortparallel }=1.4m_{0}$
and $m_{\perp }=0.36m_{0}$ ($m_{0}$ is the free electron mass), one can
estimate the spin-flip relaxation time $W_{d}^{-1}=4.7\times 10^{3}$~s. This
value exceeds the CNT relaxation time at $T=4$~K and $B=0$ by at least of
four orders of magnitude, demonstrating the significance of the radial
confinement effect in a CNT.

In conclusion, we consider electron spin relaxation in a single-walled
semiconducting CNT through the HFI with nuclear spins of $^{13}$C isotopes.
The analysis reveals the peculiarities in spin relaxation inherent to one
dimensional systems at low temperatures and/or weak magnetic fields. As a
result, it becomes dependent on the non-magnetic electron scattering.
Numerical estimations illustrate the relative importance of this relaxation
mechanism in a CNT compared to the similar processes in bulk diamond and
other carbon-based structures; strong enhancement due to the radial
confinement of electrons helps making the HFI dominant over the spin-orbital
interactions, particularly at weak magnetic fields and low temperatures.
However, the anticipated spin relaxation time of the order of 1 s in CNTs is
still much longer than those found in conventional semiconductor structures.

This work was supported in part by the SRC/MARCO Center on FENA and US Army
Research Office.

\clearpage
\begin{figure}[tbp]
\includegraphics[scale=1,angle=0]{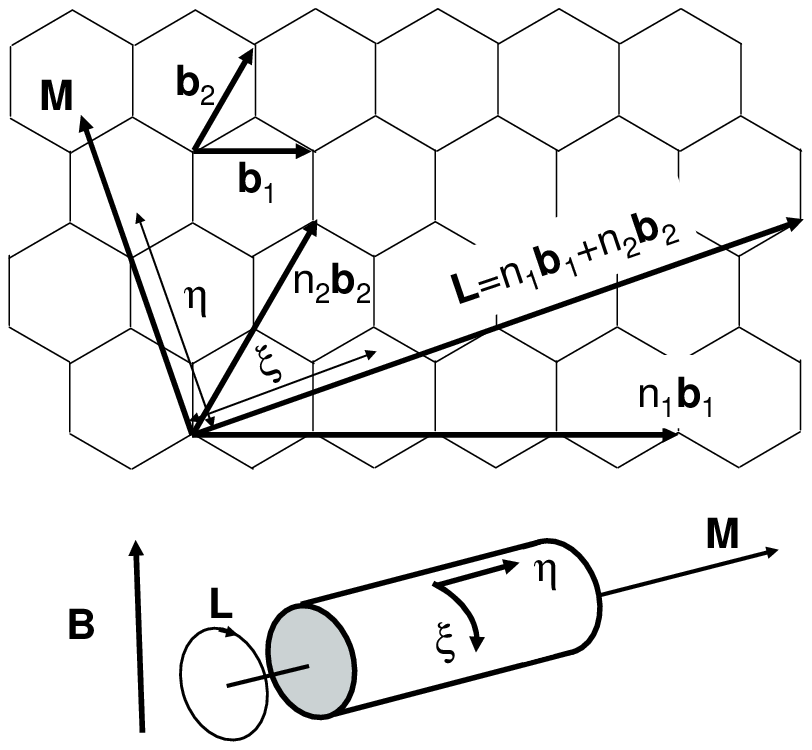}
\caption{Upper: Lattice structure of the graphene sheet. The carbon atoms
are located at the vertices of hexahedrons. $\protect\overrightarrow{b}_{1}$
and $\protect\overrightarrow{b}_{2}$ are primitive translation vectors. $%
\protect\overrightarrow{L}$ is the chiral vector and $\protect%
\overrightarrow{M}$ denotes the direction perpendicular to $\protect%
\overrightarrow{L}$. The lengths of the vectors are $b=\protect\sqrt{3}
a_{C-C}$; $L=b\protect\sqrt{n_{1}^{2}+n_{2}^{2}+n_{1}n_{2}} $; $a_{C-C}$ is
the distance between the nearest carbon atoms. The figure depicts the
particular case of $n_{1}=4$, $n_{2}=2$. Lower: CNT as a tortile graphene
sheet. $\protect\xi $ and $\protect\eta $ denote the coordinates for the
electronic states. The direction of the magnetic field $\protect%
\overrightarrow{B}$ is also shown.}
\end{figure}

\clearpage
\begin{figure}[tbp]
\includegraphics[scale=1.1]{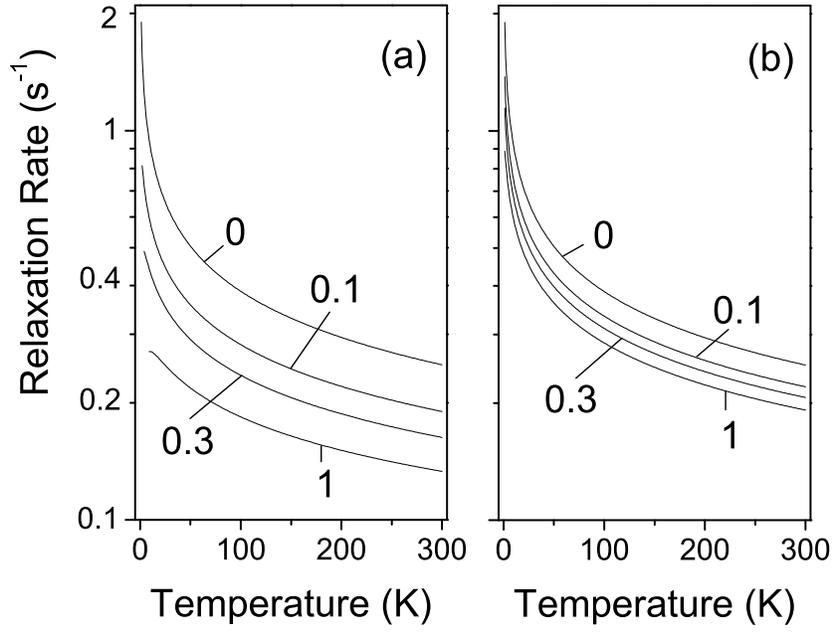}
\caption{ Calculated spin relaxation rates (a) $T_{1}^{-1}$ and (b) $%
T_{2}^{-1}$ in a (8,0) zigzag CNT as a function of temperature for different
values of magnetic field $B $. The strength of the magnetic field is
indicated in units of Tesla. }
\end{figure}

\end{document}